\documentclass[twocolumn,showpacs,amsmath,amssymb,prb]{revtex4}

\usepackage{graphicx}
\usepackage{dcolumn}
\usepackage{bm}

\begin{document}

\title{Effects of Fermi surface and superconducting gap structure in the
field-rotational experiments: A possible explanation of the cusp-like singularity in YNi$_2$B$_2$C}

\author{Masafumi Udagawa, Youichi Yanase, and Masao Ogata}

\affiliation{
Department of Physics, University of Tokyo, Hongo, Tokyo 113-0033, Japan
}

\date{\today}      

\begin{abstract}
We have studied the field-orientational dependence of zero-energy
 density of states (FODOS) for a series of systems with
 different Fermi surface and superconducting gap structures. Instead of
 phenomenological Doppler-shift method, we use an approximate analytical
 solution of Eilenberger equation together with self-consistent
 determination of order parameter and a variational treatment of vortex
 lattice. First, we compare zero-energy density of states (ZEDOS)
 when a magnetic field is applied in the nodal direction
 ($\nu_{node}(0)$) and in the antinodal direction ($\nu_{anti}(0)$), by
 taking account of the field-angle dependence of order parameter. As a result, we
 found that there exists a crossover magnetic field $H^*$ so that
 $\nu_{anti}(0) > \nu_{node}(0)$ for $H < H^*$, while $\nu_{node}(0) > \nu_{anti}(0)$ for  $H > H^*$,
 consistent with our previous analyses. Next, we showed that $H^*$ and the shape of
 FODOS are determined by contribution from the small part of Fermi
 surface where Fermi velocity is parallel to field-rotational plane. In
 particular, we found that $H^*$ is lowered and FODOS has broader
 minima, when a superconducting gap has point nodes, in contrast to
 the result of the Doppler-shift method. We also studied the effects of
 in-plane anisotropy of Fermi surface. We found that in-plane anisotropy of
 quasi-two dimensional Fermi surface sometimes becomes larger than the effects of Doppler-shift and can
 destroy the Doppler-shift predominant region. In particular, this
 tendency is strong in a multi-band system where superconducting
 coherence lengths are isotropic. Finally, we addressed the
 problem of cusp-like singularity in YNi$_2$B$_2$C and present a possible
 explanation of this phenomenon.

\end{abstract}
\maketitle

\section{Introduction}
While it is accepted that electron correlation drives non-s-wave-pairing
superconductivity, determination of detailed superconducting gap
structure still remains a challenge for both theory and experiment.
On a theoretical side, analyses starting at microscopic Hamiltonians
have succeeded in identifying gap structures of several materials.
On an experimental side, it is getting recognized that
field-orientational dependence of specific heat or thermal conductivity
carries a considerable information on gap structure. Since the
low-energy density of states changes according to the angle between
gap nodes and magnetic field, one can derive gap structure by tracing
a change of thermodynamic quantities as sweeping magnetic field in
several directions. Actually, this method has been applied to a series of
superconductors including Sr$_2$RuO$_4$\cite{izawa,tanatar,deguchi1,deguchi2},
CeCoIn$_5$\cite{izawa2,aoki}, $\kappa-$(ET)$_2$Cu(NCS)$_2$\cite{izawa3},
YNi$_2$B$_2$C\cite{izawa4,park}, PrOs$_4$Sb$_{12}$\cite{izawa5} and
UPd$_2$Al$_3$\cite{watanabe}. However, their results left serious
problems yet to be overcome. For example, in the case of Sr$_2$RuO$_4$,
thermal conductivity ($\kappa$) does not show any change with rotating
magnetic field within the ab-plane, while specific heat ($C$) shows clear
four-fold oscillation. As to CeCoIn$_5$, $\kappa$ and $C$ show
incompatible behavior. In YNi$_2$B$_2$C, both $\kappa$ and $C$ show a
cusp-like singularity as sweeping magnetic field within the
superconducting plane.

In the previous paper, we addressed the problem of Sr$_2$RuO$_4$
in terms of the fine-structure of density of
states\cite{udagawa}. Whereas, the latter two problems are left
unsolved. While it is proposed that the cusp-like singularity in
YNi$_2$B$_2$C is caused by point node in gap structure\cite{thalmeier},
their discussion is phenomenological in nature and unreliable, as we
will show.

The confusion as to the experiment is probably attributed to a lack of
firm theoretical basis. So far, the experimental data has been usually
interpreted in terms of the ``Doppler-shift criterion''. This
criterion says {\it larger density of states is obtained when a magnetic
field is applied in antinodal rather than nodal directions.} However,
this criterion is obtained on the basis of the phenomenological Doppler-shift method
that can be justified only at a limit of low-magnetic field. 
Hence, it is not clear whether this criterion is still valid in the
experimentally accessible higher magnetic fields.

In fact, in the previous paper, we showed that there occurs a crossover
from Doppler-shift predominant region at low magnetic fields to core
states predominant region at high magnetic fields\cite{udagawa}. In the core
state predominant region, larger density of states is obtained for field in nodal
direction, in contrast to the ``Doppler-shift criterion''. Hence, in order to derive a
correct nodal structure from the field-rotational experiments, it is
highly desirable to establish a theoretical analysis applicable more widely.

To this end, we present a more quantitative analyses of the crossover
behavior by taking account of the field-angle dependence of order parameter. In
particular, we are interested in how the Fermi surface or superconducting
gap structures affect the crossover behavior.

Furthermore, we would like to investigate field-orientational dependence of
zero-energy density of states (FODOS) in detail at the Doppler-shift
predominant region. It is interesting to clarify how information on nodal
structure is obtained from the shape of FODOS.

The structure of this paper is as follows. In the next section, we
summarize our method. In section \ref{isotropic}, we present a
quantitative study of the crossover behavior for several models with
different Fermi surface and nodal structures. In section \ref{inplane},
we examine the effects of in-plane anisotropy of Fermi surface to the
crossover behavior. In section \ref{carbide} we address the problem of
cusp-singularity in YNi$_2$B$_2$C. And section \ref{conclusion} is
devoted to conclusions.

\section{Method}
\label{method}
In this section, we introduce an approximate analytical solution of
the quasiclassical Eilenberger equation, which was originally proposed
by Pesch\cite{pesch}, and recently extended to unconventional
superconductors by Dahm et al.\cite{dahm}. With this
method, one can obtain exact results at $H=H_{c2}$, and quantitatively
reliable results to much lower fields. All our analyses in this
paper are based on this approximate analytical approach.

We start with the quasiclassical Eilenberger
equation\cite{eilenberger,larkin}. This equation provides a convenient
method to analyze inhomogeneous state of superconductors by decoupling
slowly varying order parameter from fast oscillating Fermi-particle
degree of freedom. In this paper, we limit ourselves to the case of
spin-singlet superconductors in clean limit. Under these conditions, the
quasiclassical Eilenberger equations read
\begin{eqnarray}
\mathbf{v}_F\cdot\bigl(\nabla - i\frac{2e}{c}\mathbf{A}(\mathbf{r})\bigr)f + 2\omega_n f - 2i\Delta g = 0, \label{eilen1}
\end{eqnarray}
\begin{eqnarray}
\mathbf{v}_F\cdot\bigl(\nabla + i\frac{2e}{c}\mathbf{A}(\mathbf{r})\bigr)f^{\dagger} - 2\omega_n f^{\dagger} + 2i\Delta^* g = 0, \label{eilen2}
\end{eqnarray}
with the normalization condition
\begin{eqnarray}
g^2 - ff^{\dagger} = 1. \label{eilennorm}
\end{eqnarray}
Here, $g=g(\mathbf{r},\mathbf{k}_F,i\omega_n)$,
$f=f(\mathbf{r},\mathbf{k}_F,i\omega_n)$ and 
$f^{\dagger}=f^{\dagger}(\mathbf{r},\mathbf{k}_F,i\omega_n)$ are normal
and anomalous components of quasiclassical Green function, $\omega_n =
(2n+1)\pi T$ is Matsubara frequency, and $\Delta=\Delta
(\mathbf{r},\mathbf{k}_F)$ denotes order parameter. The order parameter
$\Delta$ is determined in a self-consistent way, by solving the
following gap equation.
\begin{eqnarray}
\Delta (\mathbf{r},\mathbf{k}_F) = -2\pi iT\nu_n(0)\sum\limits_{0<\omega_n<\omega_c}<V(\mathbf{k}_F, \mathbf{k}_F')\nonumber\\ 
\times f(\mathbf{r},\mathbf{k}_F',i\omega_n)>_{FS(\mathbf{k}_F')}. \label{gapeq}
\end{eqnarray}
Here, $\nu_n(0)=\int\limits_{FS} d\Omega\frac{1}{v_F}$ is Zero-energy
density of states (ZEDOS) in
the normal state, $\omega_c$, a cut-off energy, $V(\mathbf{k}_F,
\mathbf{k}_F')$, a paring potential. and $<\cdots>_{FS}$ means taking
average on the Fermi surface, namely, $<\cdots>_{FS}=\int\limits_{FS}
\frac{d\Omega}{\nu_n(0)}\frac{1}{v_F}\cdots$.
 
Free energy $\Omega_s$ can also be expressed with the quasiclassical
Green function in a closed form\cite{eilenberger,klein}.
\begin{eqnarray}
\Omega_s - \Omega_n = \int\limits d\mathbf{r} \frac{(\mathbf{B}-\mathbf{H})^2}{8\pi} + \pi iT\nu_n(0)\nonumber\\ \times\sum\limits_{0<\omega_n<\omega_c}<\frac{g+1}{g-1}(\Delta^*f+\Delta f^{\dagger})>_{FS}, \label{free}
\end{eqnarray}
with $\Omega_n$, free energy in the normal state, $\mathbf{H}$, external magnetic
field, and $\mathbf{B}$, total magnetic field. Equations
(\ref{eilen1})-(\ref{free}) form a basis to study inhomogeneous superconductors. 

Next, we introduce an approximate analytical solution of Eilenberger
equation. Equations (\ref{eilen1})-(\ref{eilennorm}) can be solved
analytically only at $H=H_{c2}$. However, by extending the analytical solution to lower
fields, one can obtain quantitatively reliable results for density of
states etc, well below $H_{c2}$. To obtain approximate analytical
solution at arbitrary fields, we have to make three assumptions which
are exactly justified at $H=H_{c2}$. 

First, we decouple the order parameter into the spatial and momentum
parts
\begin{eqnarray}
\Delta (\mathbf{r},\mathbf{k}_F) = \Delta_0\Psi(\mathbf{r})\Phi(\mathbf{k}_F),
\end{eqnarray}
and normalize $\Psi(\mathbf{r})$ and $\Phi(\mathbf{k}_F)$ as
\begin{eqnarray}
<|\Psi(\mathbf{r})|^2>_{\mathbf{r}} = 1,
\end{eqnarray}
\begin{eqnarray}
<|\Phi(\mathbf{k}_F)|^2>_{FS} = 1.
\end{eqnarray}
Here, $<\cdots>_{\mathbf{r}}$ means averaging in the real space. If
$\Psi(\mathbf{r})$ is periodic, this is equivalent to taking average over
a unit cell. Then, we assume the spatial variation $\Psi(\mathbf{r})$
is described by Abrikosov vortex function, namely,
\begin{eqnarray}
\Psi(\mathbf{r}) = \Psi_A(\mathbf{r}) = \frac{1}{N}\sum\limits_{n=-\infty}^{\infty}\exp\Bigl[2\pi in\frac{y-\frac{n+1}{2}y_1}{y_0}- \nonumber\\ \frac{\pi\delta}{\Lambda}(x-nx_0)^2\Bigr]. \label{abrikosov}
\end{eqnarray}
$\Psi_A(\mathbf{r})$ can be obtained by solving linearized Ginzburg-Landau
equation with different coherence lengths in x ($\xi_x$) and
in y ($\xi_y$) directions.
Here, we set $x$ and $y$ axes perpendicular to the magnetic
field. $N=(\frac{y_0}{2x_0\delta})^{\frac{1}{4}}$ is a normalization constant to make
$<|\Psi(\mathbf{r})|^2>_{\mathbf{r}} = 1$. Area
of unit cell, $\Lambda$, can be related to the total magnetic field $B$ by
\begin{eqnarray}
\Lambda = x_0y_0 = \frac{\pi c}{eB}.
\end{eqnarray}
$\delta$ denotes the anisotropy of superconducting coherence lengths,
i.e., $\delta = \frac{\xi_y}{\xi_x}$. One can estimate $\delta$ to be
$\delta\sim\sqrt{\frac{<v_{Fy}^2>}{<v_{Fx}^2>}}$ for single-band
superconductors.

Here we note that
$\Psi(\mathbf{r})$ denotes an Abrikosov vortex lattice which is spanned
by two lattice vectors, $(0,y_0)$ and $(x_0,y_1)$. When $y_1=0$,
$\Psi(\mathbf{r})$ describes a square (rectangular) lattice, while when
$y_1=\frac{y_0}{2}$, $\Psi(\mathbf{r})$ describes a triangular
lattice. However, the shape of the unit cell is unimportant in the
subsequent analyses.

As a second assumption, we set Ginzburg-Landau parameter $\kappa$ to be
much larger than 1, hence $\mathbf{B}=\mathbf{H}$. This assumption is
justified for a lot of unconventional superconductors including YNi$_2$B$_2$C.

The third assumption is that we ignore spatial fluctuation of
$g(\mathbf{r},\mathbf{k}_F,i\omega_n)$, namely, $<g^2>_{\mathbf{r}} =
<g>_{\mathbf{r}}^2$. Since at $H=H_{c2}$, the order parameter vanishes and
translational symmetry is recovered, this assumption is also justified.
The validity of this assumption has been justified for lower fields\cite{dahm}.

With these three assumptions, one can obtain an approximate analytical
solution of eqs. (\ref{eilen1})-(\ref{eilennorm}) at arbitrary magnetic
fields by utilizing the operator techniques introduced in ref. \onlinecite{schopohl}. We
summarize the basic results below. For details, see, for example,
refs. \onlinecite{dahm} and \onlinecite{graser}.

\begin{enumerate}
\item[[1]] spatial average of $g(\mathbf{r},\mathbf{k}_F,i\omega_n)$
\begin{eqnarray}
<g(\mathbf{r},\mathbf{k}_F,i\omega_n)>_{\mathbf{r}} = -\mathrm{Re}\frac{1}{\sqrt{1 + P(\mathbf{v}_F,i\omega_n)}}, \label{pgreen}
\end{eqnarray}
where,
\begin{eqnarray}
P(\mathbf{v}_F,i\omega_n) = \frac{4\Lambda}{\pi}|\Delta_0|^2\bigl(\frac{|\Phi(\mathbf{k}_F)|}{|\mathbf{v}_{F\perp}|}\bigr)^2(1\nonumber\\ 
- F(\Lambda, \mathbf{v}_{F\perp}, \omega_n)),
\end{eqnarray}
\begin{eqnarray}
F(\Lambda, \mathbf{v}_{F\perp}, \omega_n) = \int\limits_0^{\infty}du \exp\bigl(-u-\frac{\pi |\mathbf{v}_{F\perp}|^2}{8\Lambda\omega_n^2}u^2\bigr).
\end{eqnarray}
Here, effective Fermi velocity $\mathbf{v}_{F\perp}$ is expressed as
\begin{eqnarray}
\mathbf{v}_{F\perp} = \sqrt{\delta}v_{Fx}\mathbf{e}_x + \frac{1}{\sqrt{\delta}}v_{Fy}\mathbf{e}_y, \label{fermi}
\end{eqnarray}
with the projection of Fermi velocity on x ($v_{Fx}$) and y ($v_{Fy}$) axes.

\item[[2]] DOS $\nu(\epsilon)$
\begin{eqnarray}
\frac{\nu(\epsilon)}{\nu_n(0)} = -< <g(\mathbf{r},\mathbf{k}_F,i\omega_n\rightarrow\epsilon+i0)>_{\mathbf{r}} >_{FS}. \label{pdos}
\end{eqnarray}
In particular, the formula of ZEDOS becomes quite simple.
\begin{eqnarray}
\frac{\nu(0)}{\nu_n(0)} = <\frac{1}{\sqrt{1+\frac{4\Lambda}{\pi}|\Delta_0|^2\bigl(\frac{|\Phi(\mathbf{k}_F)|}{|\mathbf{v}_{F\perp}|}\bigr)^2}}>_{FS}. \label{pzedos}
\end{eqnarray}

\item[[3]] gap equation

In order to obtain gap equation, we decouple $V(\mathbf{k}_F, \mathbf{k}_F')$ as $V(\mathbf{k}_F, \mathbf{k}_F') = V\Phi^*(\mathbf{k}_F)\Phi(\mathbf{k}_F')$, then we obtain
\begin{eqnarray}
\log(\frac{2e^{\gamma}\omega_c}{\pi T_c}) = \sum\limits_{0<\omega_n<\omega_c}\frac{2\pi T_c}{\omega_n}<|\Phi(\mathbf{k}_F)|^2\nonumber\\ 
\times\frac{F(\Lambda, \mathbf{v}_{F\perp}, \omega_n)}{\sqrt{1 + P(\mathbf{v}_F,i\omega_n)}}>_{FS}, \label{peschgapeq}
\end{eqnarray}
where $\gamma = 0.5772\cdots$ is Euler's constant. The formula to
determine $H_{c2}$ can be obtained by setting $\Delta_0 = 0$ in eq. (\ref{peschgapeq}).
\begin{eqnarray}
\log(\frac{2e^{\gamma}\omega_c}{\pi T_c}) = \int\limits_0^{\infty}ds\frac{1-\exp (-\frac{\omega_cs}{\pi T})}{\sinh s}<|\Phi(\mathbf{k}_F)|^2\nonumber\\ 
\times e^{-\frac{|\mathbf{v}_{F\perp}|^2}{8\pi T^2\Lambda}s^2}>_{FS}. \label{phc2}
\end{eqnarray}

\item[[4]] free energy
\begin{eqnarray}
\Omega_s-\Omega_n = -\nu_n(0)|\Delta_0|^2\sum\limits_{0<\omega_n<\omega_c}\frac{2\pi T_c}{\omega_n}<|\Phi(\mathbf{k}_F)|^2 \nonumber\\ 
\times\frac{P(\mathbf{v}_F,i\omega_n)F(\Lambda, \mathbf{v}_{F\perp}, \omega_n)}{\sqrt{1+P(\mathbf{v}_F,i\omega_n)}(1+\sqrt{1+P(\mathbf{v}_F,i\omega_n)})^2}>_{FS}. \label{pfree}
\end{eqnarray}

\end{enumerate}
We calculate density of states by the following procedure. First,
Fermi velocity $\mathbf{v}_F$ and nodal structure $\Phi(\mathbf{k}_F)$
are determined by assuming a phenomenological model or starting at a
microscopic Hamiltonian. Then, at a given temperature $T$ and
a magnetic field $\mathbf{H}$, one can calculate
order parameter amplitude $\Delta_0$ by solving
eq. (\ref{peschgapeq}). By substituting
$\Delta_0$ into eq. (\ref{pdos}), one can obtain DOS
$\nu(\epsilon)$. When magnetic field is applied off a symmetry axis,
the ratio of coherence lengths $\delta$ deviates from 1. In this case,
we determine $\delta$ variationally by minimizing free energy
eq. (\ref{pfree}) with respect to $\delta$. This prescription is
equivalent to the variational ansatz of Abrikosov vortex lattice
introduced in refs. \onlinecite{graser} and \onlinecite{dahm2}.

One can incorporate multi-band properties into this approximate
analytical approach with a slight modification of [1]-[4]\cite{graser,kusunose}.
In multi-band systems, $\mathbf{k}_F$, $\mathbf{v}_F$, $f$,
$f^{\dagger}$, $g$, $\Delta$, $\nu_n(0)$ become band-dependent. To
specify band index, we write these quantities as $\mathbf{k}_F^{\alpha}$,
$\mathbf{v}^{\alpha}_F$, $f^{\alpha}$, $f^{\alpha\dagger}$,
$g^{\alpha}$, $\Delta^{\alpha}$, $\nu_n^{\alpha}(0)$. Next, we decouple
order parameter as
\begin{eqnarray}
\Delta^{\alpha}(\mathbf{r},\mathbf{k}_F^{\alpha}) = \Delta_0^{\alpha}\Psi^{\alpha}(\mathbf{r})\Phi^{\alpha}(\mathbf{k}_F^{\alpha}). \label{morder}
\end{eqnarray}
Here, we assume $\Psi^{\alpha}(\mathbf{r})$ to be described by
Abrikosov lattice, $\Psi_A(\mathbf{r})$. Furthermore, we assume that the
anisotropy of coherence lengths $\delta^{\alpha}$ are the same for all
bands ($\delta^{\alpha}=\delta$), and determine $\delta$ by minimizing
free energy. Due to inter-band mixing effect by a pair-hopping term, the spatial
variation of $\Delta$ of each bands are not independent any longer.
Then, one can take account of the inter-band mixing effect
through minimization of $\delta$.

As we will discuss in section. \ref{carbide}, we are interested in 
a system where quasi-two-dimensional Fermi surface and three-dimensional
Fermi surface co-exist. In such a system, superconducting coherence
length may be isotropic due to the three-dimensional Fermi
surface, and lead to isotropic vortex structure.
In this case, one may expect unusual phenomena characteristic of
multi-band system, in the sense that quasi-two-dimensional Fermi surface
and isotropic vortex structure are incompatible in a single-band superconductor.

Here, we note that some complication may arise in vortex
structure due to multi-band effects. For example, it is proposed that
order parameter of different bands may have zero-points at different
places. However, we believe our results will not be influenced by the
detail of vortex structure and ignore them.

Here, let us check how [1]-[4] are modified in a multi-band case.
[1]spatial average of $g(\mathbf{r},\mathbf{k}_F,i\omega_n)$ also
holds for multi-band systems, by making each quantity in
eqs. (\ref{pgreen})-(\ref{fermi}) band-dependent.
[2]DOS $\nu(\epsilon)$ and [4]free energy are also correct by
making the quantities band-dependent, and summing over band indices in
the right sides of eqs. (\ref{pdos}), (\ref{pzedos}), and (\ref{pfree}).
Only gap equation in [3] needs substantial modification. In a multi-band
system, we decouple pair
potential $V^{\alpha\beta}(\mathbf{k}_F^{\alpha},
\mathbf{k}_F'^{\beta})$ as $V(\mathbf{k}_F^{\alpha},
\mathbf{k}_F'^{\beta}) =
V^{\alpha\beta}\Phi^*(\mathbf{k}_F^{\alpha})\Phi(\mathbf{k}_F'^{\beta})$,
then gap equation eq. (\ref{peschgapeq}) becomes
\begin{eqnarray}
\Delta_0^{\alpha}\log(\frac{2e^{\gamma}\omega_c}{\pi T_c}) = \sum\limits_{\beta}\nu_n^{\beta}(0)V^{\alpha\beta}\Delta_0^{\beta}\sum\limits_{0<\omega_n<\omega_c}\frac{2\pi T_c}{\omega_n}\nonumber\\ 
\times <|\Phi^{\beta}(\mathbf{k}_F^{\beta})|^2\frac{F^{\beta}(\Lambda, \mathbf{v}_{F\perp}^{\beta}, \omega_n)}{\sqrt{1 + P^{\beta}(\mathbf{v}_F^{\beta},i\omega_n)}}>_{FS(\mathbf{k}_F^{\beta})}, \label{mpeschgapeq}
\end{eqnarray}

\section{Quantitative study of the crossover behavior}
\label{isotropic}
In this section, we present a quantitative study of the crossover behavior
for the following three models. Below, we take $\mathbf{e}_a$, $\mathbf{e}_b$ and
$\mathbf{e}_c$ as a right-handed orthonormal basis.
\begin{enumerate}
\item[(a)] line nodes on spherical Fermi surface 
\begin{eqnarray}
\mathbf{v}_F = v_F(\sin\theta\cos\phi\mathbf{e}_a + \sin\theta\sin\phi\mathbf{e}_b + \cos\theta\mathbf{e}_c).
\end{eqnarray}
\begin{eqnarray}
\Phi(\mathbf{k}_F) = \sqrt{\frac{15}{4}}\sin^2\theta\sin 2\phi.
\end{eqnarray}

\item[(b)] line nodes on cylindrical Fermi surface 
\begin{eqnarray}
\mathbf{v}_F = v_F(\cos\phi\mathbf{e}_a + \sin\phi\mathbf{e}_b + \epsilon\sin k_z\mathbf{e}_c).
\end{eqnarray}
\begin{eqnarray}
\Phi(\mathbf{k}_F) = \sqrt{2}\sin 2\phi.
\end{eqnarray}
Here, $\epsilon$ gives a small c-axis dispersion. We set
$\epsilon = 0.05$.

\item[(c)] point nodes on spherical Fermi surface
\begin{eqnarray}
\mathbf{v}_F = v_F(\sin\theta\cos\phi\mathbf{e}_a + \sin\theta\sin\phi\mathbf{e}_b + \cos\theta\mathbf{e}_c).
\end{eqnarray}
\begin{eqnarray}
\Phi(\mathbf{k}_F) = \sqrt{\frac{15}{19}}(1 - \sin^2\theta\cos 4\phi). \label{gap_sph_p}
\end{eqnarray}
\end{enumerate}

These models are often used to approximate more complicated realistic
materials. Hence, it is important to study the basic properties of these
models in detail.

Throughout our paper, we specify the direction of magnetic field with
two parameters $\theta_0$ and $\phi_0$ as
\begin{eqnarray}
\mathbf{H} = H(\sin\theta_0\cos\phi_0\mathbf{e}_a + \sin\theta_0\sin\phi_0\mathbf{e}_b + \cos\theta_0\mathbf{e}_c).
\end{eqnarray}
In principle, we fix $\theta_0 = 90^{\circ}$ and study the change of
ZEDOS with varying $\phi_0$ in this paper. In other words, we consider
the ab-plane as a field-rotational plane. As to the three models above,
$\phi_0=0^{\circ}$ and $\phi_0=45^{\circ}$ correspond to the cases where
a magnetic field is applied in nodal direction and antinodal direction,
respectively.

Here, we focus on ZEDOS at low temperatures, because universal behaviors
are expected in this region. We note that some cares are necessary to
discuss thermodynamic quantities at higher temperature, because DOS at
high energies depend on the magnetic field direction quite differently
from ZEDOS\cite{udagawa}. Thus we fix $T = 0.1T_c$ and $\omega_c = 20\pi
T_c$ throughout this section. At this temperature, ZEDOS well represents
the thermodynamic properties of a system.

\subsection{Crossover behavior of ZEDOS and order parameter}
First, we consider ZEDOS when magnetic field is applied in the nodal
direction ($\nu_{node}(0)$) and in the antinodal direction
($\nu_{anti}(0)$). In the previous paper\cite{udagawa}, we showed that
the Doppler-shift criterion is correct at low fields, while it is broken
at high fields due to the quasiparticles propagating parallel to
magnetic field. In the previous paper, we used an assumption that
$\Delta_0$ does not depend on field direction. In the present paper, we calculate ZEDOS
with determining $\Delta_0$ self-consistently. In general, $\Delta_0$
changes as field direction varies. In particular, near $H_{c2}$, in-plane
anisotropy of $H_{c2}$ influences FODOS through the field-angle dependence of
$\Delta_0$. Perhaps, the anisotropy of $H_{c2}$ continues to affect FODOS
to lower fields, and suppress the Doppler-shift predominant
region. Hence, it is of interest to estimate how the field-angle
dependence of $\Delta_0$ modifies the crossover behavior.

\begin{figure}
\includegraphics[width=0.5\textwidth]{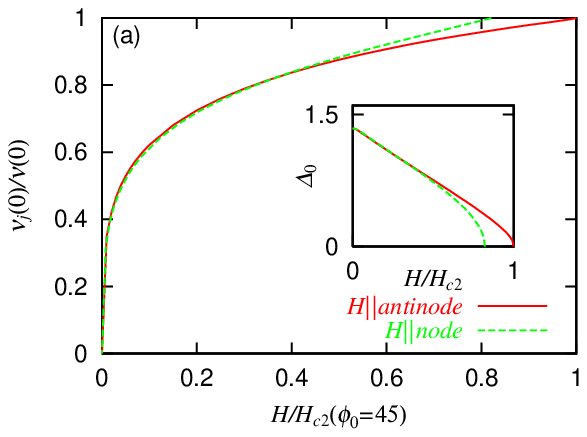}
\includegraphics[width=0.5\textwidth]{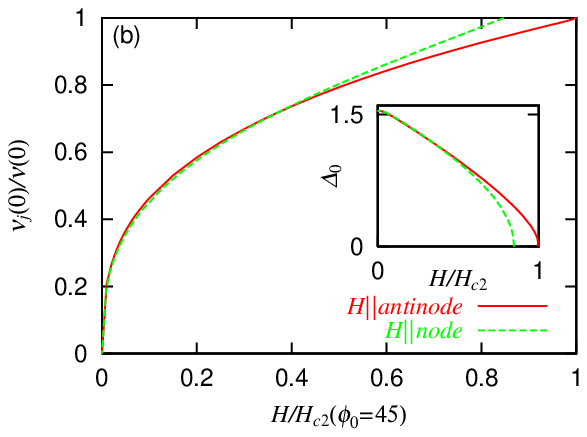}
\includegraphics[width=0.5\textwidth]{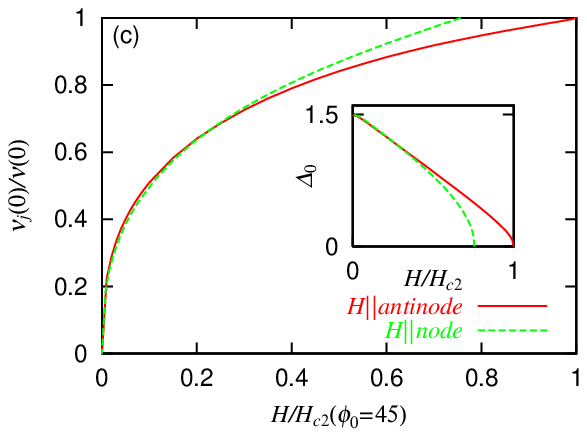}
\caption{\label{f_fodosamp}Field dependence of ZEDOS under magnetic
 field in the nodal and antinodal directions for (a)line nodes on spherical Fermi
 surface, (b)line nodes on cylindrical Fermi surface,
 and (c)point nodes on spherical Fermi surface. (Inset) Field dependence
 of $\Delta_0$. $\Delta_0$ is normalized with T$_c$, and magnetic field is normalized
 with $H_{c2}$ when a field is in the antinodal direction. The agreement
 between (a) and the result of Eilenberger approach (fig. 4 of
 ref. \onlinecite{miranovic}) is quite good.}
\end{figure}

\begin{figure}
\includegraphics[width=0.5\textwidth]{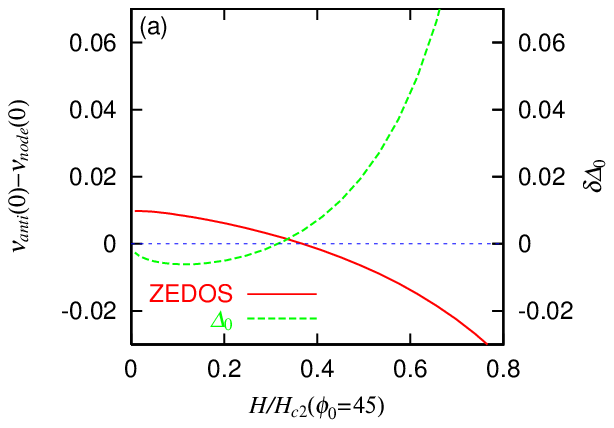}
\includegraphics[width=0.5\textwidth]{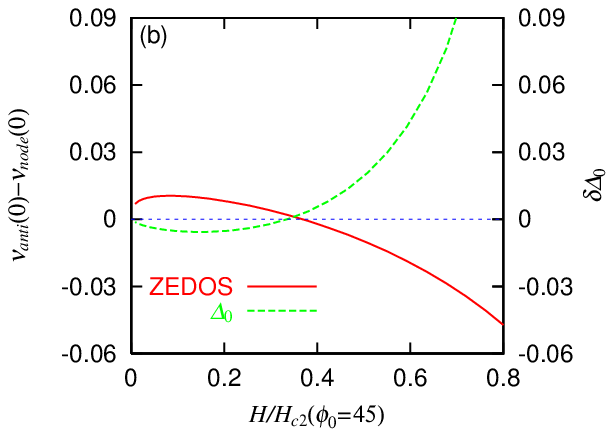}
\includegraphics[width=0.5\textwidth]{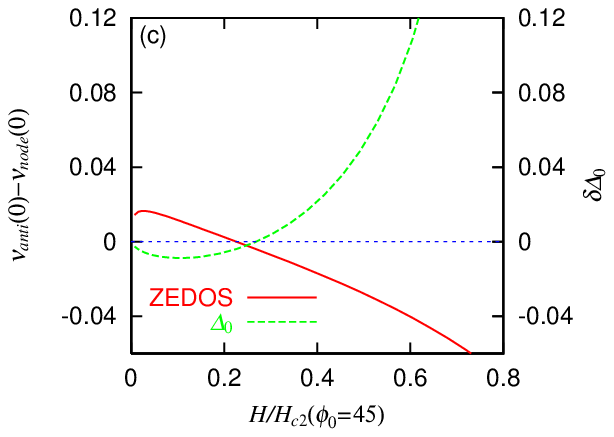}
\caption{\label{diff}Field dependence of $\delta\Delta_0$ and
 $\nu_{anti}(0)-\nu_{node}(0)$ for (a)line nodes on spherical Fermi
 surface, (b)line nodes on cylindrical Fermi surface,
 and (c)point nodes on spherical Fermi surface. $\delta\Delta_0$ is
 normalized with T$_c$, and magnetic field is normalized
 with $H_{c2}$ when a field is in the antinodal direction.}
\end{figure}

In fig. \ref{f_fodosamp}, we show the FODOS for the three models introduced above.
These show a similar crossover behavior with
different crossover fields $H^*$. For $H < H^*$, $\nu_{anti}(0) >
\nu_{node}(0)$, consistent with the ``Doppler-shift criterion'', while
for $H > H^*$, $\nu_{node}(0) > \nu_{anti}(0)$. These results are
consistent with our previous analyses\cite{udagawa}. Since $H_{c2}$ is smaller when
field is in the nodal direction ($H\parallel n$) than in the antinodal
direction ($H\parallel a$) in all the three cases, the anisotropy of
$H_{c2}$ generally supports the relation $\nu_{node}(0) > \nu_{anti}(0)$ at high fields.

In order to examine how the anisotropy of $\Delta_0$ affects FODOS at
low fields, we plot $\delta\Delta_0\equiv\Delta_0(H\parallel
a)-\Delta_0(H\parallel n)$ in fig. \ref{diff}, together with the anisotropy of
ZEDOS, $\nu_{anti}(0)-\nu_{node}(0)$. One can find that $\delta\Delta_0$
changes its sign near $H^*$ and becomes negative for $H \lesssim
H^*$. This means that anisotropy of $\Delta_0$ has the same sign with
the relation $\nu_{anti}(0) > \nu_{node}(0)$ in the Doppler-shift
predominant region. Combining this with the effects of anisotropic $H_{c2}$ at
high magnetic fields, it is shown that our previous result\cite{udagawa} on the crossover
behavior is not altered by taking account of the field-angle dependence of $\Delta_0$.

Next, let us study individual cases. First, let us compare the cases of line
nodes on spherical Fermi surface(fig. \ref{diff}(a)) and on
cylindrical Fermi surface(fig. \ref{diff}(b)). In
spite of the difference in dimensionality of Fermi surfaces, both
cases show similar behavior of $\nu_{node}(0)$ and $\nu_{anti}(0)$ with
very close $H^*$. In fact, the small c-axis
dispersion of the cylindrical Fermi surface plays a key role in this
similarity. In eq. (\ref{pzedos}), information of Fermi surface is
included only in $|\mathbf{v}_{F\perp}|$. In the case of cylindrical
Fermi surface, $|\mathbf{v}_{F\perp}|^2\propto\delta\cos^2(\phi - \phi_0) +
\frac{1}{\delta}\epsilon^2\sin^2k_z$. While $\epsilon$ is much smaller than 1, one
cannot neglect the term $\frac{1}{\delta}\epsilon^2\sin^2k_z$.
In the case of anisotropic Fermi surface, one has $\delta =
\frac{\xi_y}{\xi_x}\sim\sqrt{\frac{<v_c^2>}{<v_{ab}^2>}}\sim\epsilon$ in
the whole range of magnetic field, as a result of minimizing the free
energy eq. (\ref{free}). Hence, the in-plane and the c-axis components
of effective Fermi velocity, $\delta\cos^2(\phi - \phi_0)$ and $\frac{1}{\delta}\epsilon^2\sin^2k_z$
are scaled to be equivalent in magnitude. 

Next, we study the case of point nodes on spherical Fermi surface
(fig. \ref{diff}(c)). In this case, one can find that the crossover field
$H^*$ is smaller ($H^*\sim 0.2H_{c2}$) than the case of line nodes
($H^*\sim 0.4H_{c2}$). This difference in $H^*$ can be understood in terms of
the shape of gap minima. To see this, we note that
the reversal of the Doppler-shift criterion at high magnetic fields is
caused by the quasiparticles propagating parallel to magnetic field\cite{udagawa}. At high magnetic fields,
contribution from such quasiparticles to ZEDOS, $\nu_{QP\parallel H}$, is
smaller for $H\parallel a$ than $H\parallel n$, resulting in the
relation $\nu_{node}(0) > \nu_{anti}(0)$. Here, for the point node case,
$|\Phi(\mathbf{k}_F)|$ is quadratic near gap minima (for example, at
$\theta=90^{\circ},\phi=0^{\circ}$), while it is linear in the line node cases. 
Therefore, the order parameter is smaller near
gap minima in the point node case, leading to larger
$\nu_{QP\parallel H}$ for $H\parallel n$. On the other hand, there is not
much difference in $\nu_{QP\parallel H}$ when $H\parallel a$.
As a result, $\nu_{node}(0) - \nu_{anti}(0)$ is larger in the point node
case and continues to be positive to lower field.

\begin{figure}
\includegraphics[width=0.5\textwidth]{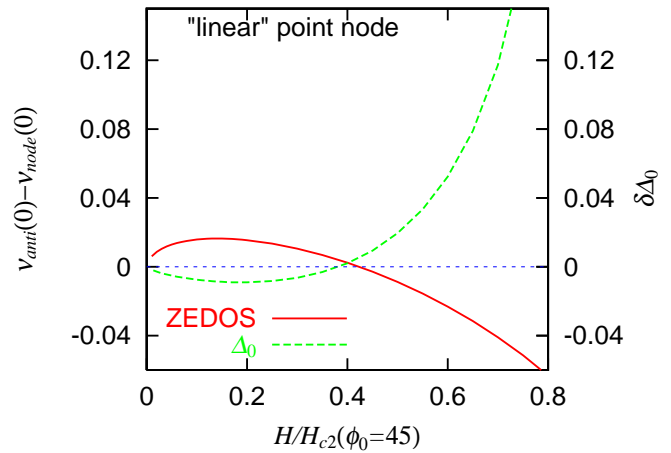}
\caption{\label{diff_sharp}Field dependence of $\delta\Delta_0$ and
 $\nu_{anti}(0)-\nu_{node}(0)$ for spherical Fermi surface with point
 nodes. Gap function is given by
 $\Phi(\mathbf{k}_F)=\sqrt{1-\sin^2\theta\cos 4\phi}$.  }
\end{figure}

One may suppose that the smaller $H^*$ can be attributed to the gap
structure in the ``polar part''($\theta\sim 0^{\circ}$). Since in the
point node case, $|\Phi(\mathbf{k}_F)|\not= 0$ except for the
``equatorial part''($\theta\sim90^{\circ}$) of the Fermi surface, the
effect of Doppler-shift is weaker in the ``polar part'', and may lead to the smaller $H^*$. However, this
does not hold. In fact, if we replace the gap function
eq. (\ref{gap_sph_p}) with $\Phi(\mathbf{k}_F)=\sqrt{1-\sin^2\theta\cos
4\phi}$, i.e., a point node model which varies linearly near gap minima, we
find that $H^*\sim 0.4H_{c2}$(fig. \ref{diff_sharp}), comparable to the line node cases,
although $|\Phi(\mathbf{k}_F)|$ is larger than the quadratic point node
model (eq. (\ref{gap_sph_p})) everywhere on the Fermi surface.

\subsection{Field-orientational dependence of ZEDOS}
\begin{figure}
\includegraphics[width=0.5\textwidth]{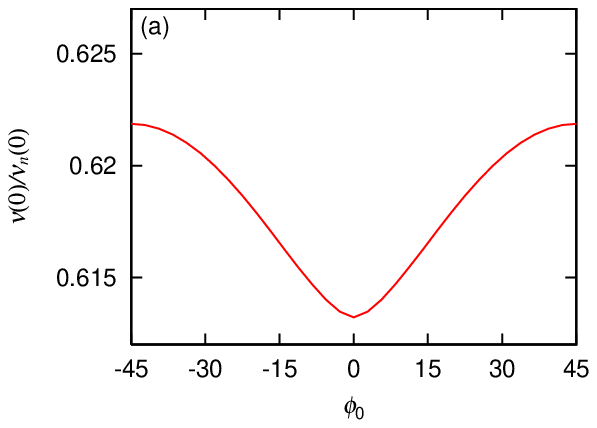}
\includegraphics[width=0.5\textwidth]{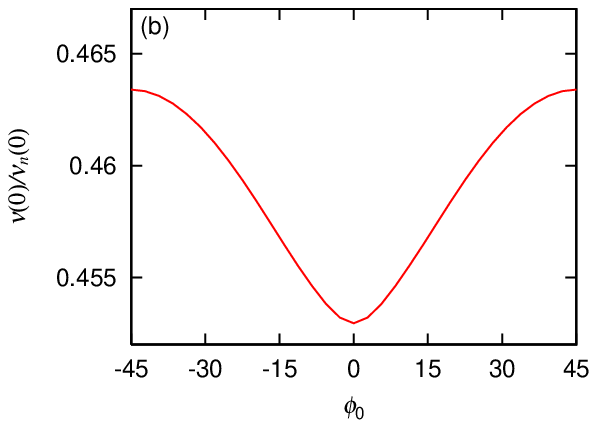}
\includegraphics[width=0.5\textwidth]{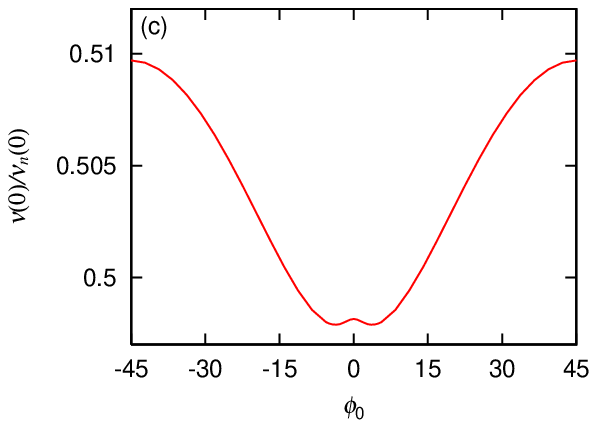}
\caption{\label{f_fodos}Field-orientational dependence of ZEDOS for
 (a)line nodes on spherical Fermi surface, (b)line nodes on cylindrical Fermi
 surface, and (c)point nodes on spherical Fermi surface.}
\end{figure}

\begin{figure}
\includegraphics[width=0.5\textwidth]{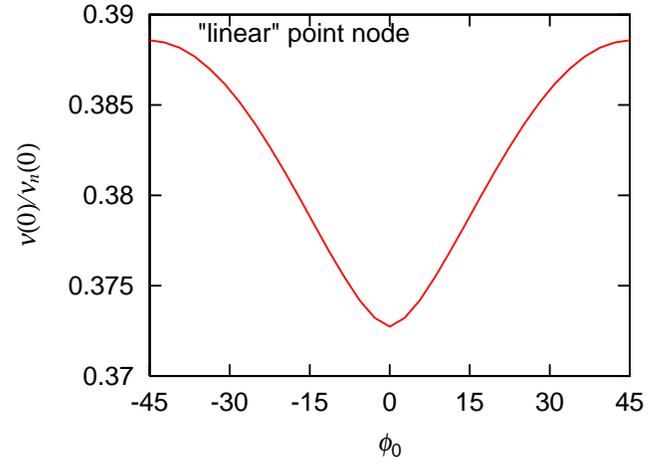}
\caption{\label{sharp_p}Field-orientational dependence of ZEDOS for
 (a)spherical Fermi surface with point nodes.  Gap function is given by
 $\Phi(\mathbf{k}_F)=\sqrt{1-\sin^2\theta\cos 4\phi}$.}
\end{figure}

In this subsection, we study not only the amplitude but also the whole
field-orientational dependence of ZEDOS. In particular, the shape of
minima is of our interest, since we will discuss the origin of cusp-like
minima which has been attributed to a point node in the Doppler-shift
method\cite{thalmeier}. In fig. \ref{f_fodos}, we show our results for
the three models. These calculations are done at $H=0.1H_{c2}$. One can
find similar FODOS for the line node cases (fig. \ref{f_fodos}(a),(b))
with a slight difference in amplitudes. Whereas in the point node case
(fig. \ref{f_fodos}(c)), one obtains FODOS with broader minima than
the line node cases. Furthermore, in the point node case, tiny maxima appear when magnetic
field is applied right in the nodal direction
($\phi_0=0^{\circ}$). These features again reflect the quadratic
variation of order parameter near the gap minima. For comparison, we show FODOS for the gap function
$\Phi(\mathbf{k}_F)=\sqrt{1-\sin^2\theta\cos 4\phi}$ in
fig. \ref{sharp_p}, where one can find narrower minima similar
to the case of line nodes. In both cases, our calculation does not
support the results of Doppler-shift method.

\begin{figure}
\includegraphics[width=0.5\textwidth]{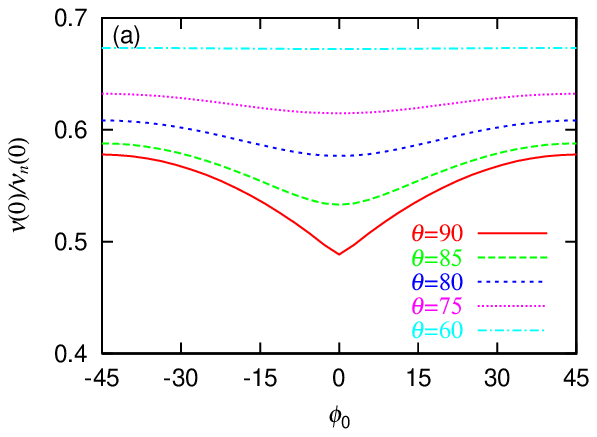}
\includegraphics[width=0.5\textwidth]{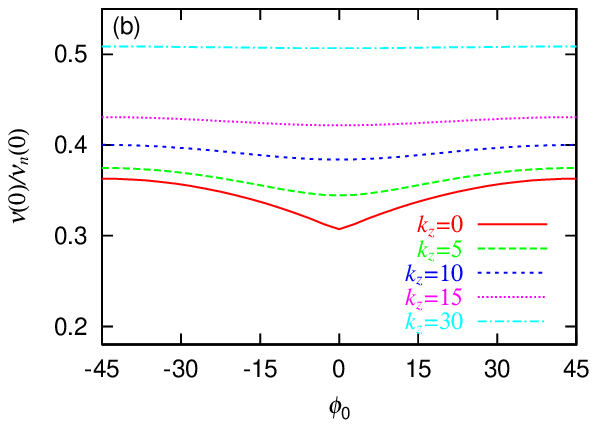}
\includegraphics[width=0.5\textwidth]{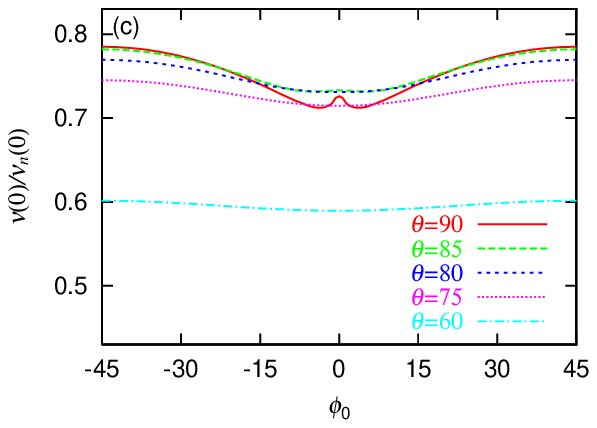}
\caption{\label{fermi_div}Contribution to FODOS from several parts of the
 Fermi surface for (a)line nodes on spherical Fermi surface, (b)line
 nodes on cylindrical Fermi surface, and(c)point nodes on spherical
 Fermi surface. In these three cases, total FODOS is given by averaging
 over $\theta$ ($0^{\circ}<\theta<90^{\circ}$) or $k_z$ ($-\pi<k_z<\pi$).}
\end{figure}

For the better understanding of the features in FODOS, we plot
contribution to FODOS from several parts of the Fermi surface in
fig. \ref{fermi_div}. From these figures, one can find that FODOS is
determined by the contribution from $\theta\sim 90^{\circ}$ on
the spherical Fermi surface or $k_z\sim 0^{\circ}$ on the cylindrical
Fermi surface. In other words, Most contribution comes from the part
where effective Fermi velocity $\mathbf{v}_{F\perp}$(eq. (\ref{fermi}))
is parallel to the field-rotational plane. One can coherently understand
the features of FODOS and $H^*$ obtained in this section from this
viewpoint.

First, the difference between point node and line node is unimportant,
because these gap structures behave similarly at $\theta\sim
90^{\circ}$, where $\mathbf{v}_{F\perp}$ is parallel to the field-rotational
plane. The dimensionality of Fermi surface is also unimportant. While
Fermi velocity is almost parallel to the field-rotational plane on a
cylindrical Fermi surface, the effective Fermi velocity is so only near
$k_z\sim 0^{\circ}$, due to small $\delta$($\sim\epsilon$). Whereas it is important whether order
parameter varies linearly or quadratically near gap minima, because this
difference lies in $\theta\sim 90^{\circ}$. In this sense, $H^*$ {\it and
FODOS are determined by the local structure of order parameter on the
part where} $\mathit{v_{F\perp}}$ {\it is parallel to the
field-rotational plane, rather than the global nodal structure of the
entire Fermi surface}.

\section{Effects of in-plane anisotropy of Fermi
 surface}
\label{inplane}
In this section, we examine the effects of in-plane anisotropy of Fermi
surface on FODOS. The anisotropy of Fermi surface affects ZEDOS through
$|\mathbf{v}_{F\perp}|$ in eq. (\ref{pzedos}). If the
distribution of $|\mathbf{v}_{F\perp}|$ changes
according to the magnetic field direction, ZEDOS also depends on the
field direction, even without gap nodes. In this section,
we focus on the shift of $H^*$ with increasing anisotropy of Fermi
surface.

\begin{figure}
\includegraphics[width=0.5\textwidth,scale=0.3]{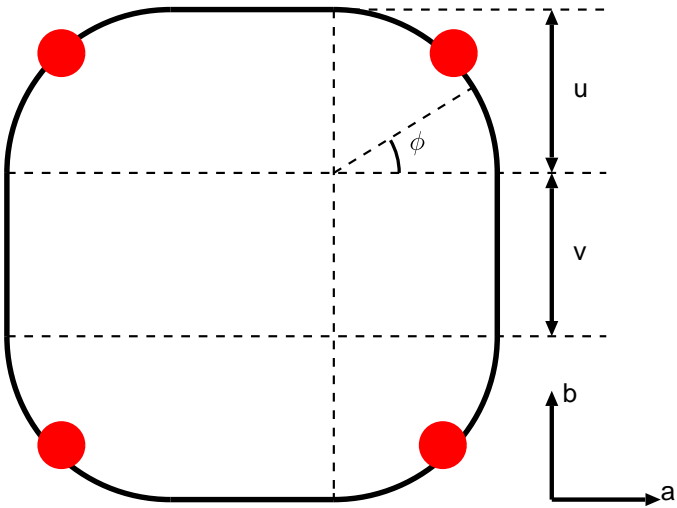}
\caption{\label{necylinder} A schematic picture of the Fermi surface with
 in-plane anisotropy seen from the c-axis. Position of gap node is shown with filled circles}
\end{figure}

In order to take account of the in-plane anisotropy of Fermi surface, we
consider a simplified model shown in fig. \ref{necylinder}. We make
Fermi surface by connecting the sets of wave numbers, i.e. by adding
nested parts to a cylindrical Fermi surface. 

\begin{eqnarray}
(k_a, k_b) = (\pm(u+v),k_b), \hspace{0.3cm} (-v<k_b<v), \label{nec1}
\end{eqnarray}
\begin{eqnarray}
(k_a, k_b) = (k_a,\pm(u+v)), \hspace{0.3cm} (-v<k_a<v), \label{nec2}
\end{eqnarray}
\begin{eqnarray}
(k_a, k_b) = (\pm v\pm u\cos\phi, \pm v\pm u\sin\phi),  \hspace{0.1cm} (0<\phi<\frac{\pi}{2}). \label{nec3}
\end{eqnarray}
Here, $\zeta=v/u$ is a measure of anisotropy. When $\zeta=0.5$ and $
1.0$, the linear parts occupy $24.1$ and $38.9\%$ of the Fermi surface, respectively.

We assume the absolute value of in-plane components of Fermi velocity $v_F$
are the same all over the Fermi surface, and set a c-axis component
$v_{Fc}=0.05v_F\sin k_z$. We assume a d$_{x^2-y^2}$-like pairing, i.e. 
\begin{eqnarray}
\Phi(\mathbf{k}_F) \propto\cos(2\tan^{-1}(\frac{k_b}{k_a})),
\end{eqnarray}
i.e., gap nodes are on the cylindrical parts of the Fermi surface.

\begin{figure}
\includegraphics[width=0.5\textwidth]{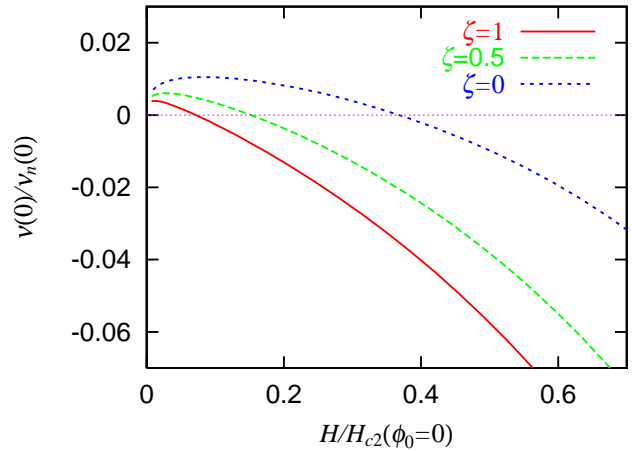}
\caption{\label{nec_zedos} Comparison of $\nu_{node}(0)$ and
$\nu_{anti}(0)$ for $\zeta=0, 0.5, 1.0$. Magnetic field is normalized
 with $H_{c2}$ when a field is in the antinodal direction.}
\end{figure}

When magnetic field is applied perpendicular to one of the nested parts,
$|\mathbf{v}_{F\perp}|\sim 0$ on the nested parts, leading to
suppression of ZEDOS. Whereas, when magnetic field is applied in the
nodal direction, the nested parts does not influence ZEDOS so
much. Hence, it is expected that the Doppler-shift predominant region is
suppressed as $\zeta$ is increased. In fig. \ref{nec_zedos}, we show
$\nu_{anti}(0)-\nu_{node}(0)$ for several values of $\zeta$.
One can find that $H^*$ is lowered as $\zeta$ is increased. At
$\zeta=1.0$, the Doppler-shift predominant region is suppressed
to $H\sim 0.05H_{c2}(\phi_0=45^{\circ})$, i.e., the almost entire field range at
which experiments have been done.

This result suggests that many cares should be taken to apply the
 ``Doppler-shift criterion'' when Fermi surface is highly anisotropic. 
At least one should check that maxima and minima
of FODOS are exchanged at high fields and low fields, in order to
confirm that the system enters the Doppler-shift predominant region at
low magnetic fields. It seems that recent experiments do not find any
 exchange of maxima and minima\cite{izawa2,aoki}.

\section{Origin of the cusp-like singularity}
\label{carbide}
In this section, we study origin of the cusp-like singularity in FODOS
observed through specific heat\cite{park} and thermal conductivity\cite{izawa4} in
YNi$_2$B$_2$C. In both cases, the
cusp-like minima are observed when magnetic field is applied parallel to
a or b-axis. So far, the cusp-like minima are attributed to point nodes
on the superconducting gap. In particular, s$+$g-wave gap structure has
been proposed for YNi$_2$B$_2$C\cite{thalmeier}. However, as we showed
in section \ref{isotropic}, rather broader minima of FODOS are obtained from
s$+$g-wave-like quadratic point nodes.

Here, we would like to propose a new mechanism that leads to cusp-like
minima in FODOS. In our theory, two conditions are required for the
appearance of cusp-like minima. The first condition is that a system has
to have a quasi-two-dimensional nested Fermi surface. And the second condition is
that the superconducting coherence lengths are isotropic. Under these
conditions, cusp-like minima appear when a magnetic field is applied
perpendicular to the nested part of the Fermi surface.

Here, we briefly describe how cusp-like minima appear from the above
conditions, in terms of eq. (\ref{pzedos}). Provided that
superconducting coherence lengths are isotropic, one obtains
$\delta=\frac{\xi_c}{\xi_a}\sim 1$. When the magnetic field is
applied at an angle $\phi_0$ with the normal line of the nested part of
the Fermi surface, we have
\begin{eqnarray}
|\mathbf{v}_{F\perp}|^2 = v_F^2(\sin^2\phi_0 + \epsilon^2\sin^2 k_z).
\end{eqnarray}
on the nested part, with a c-axis component of Fermi velocity,
 $\epsilon\sin k_z$. If the c-axis dispersion is weak
enough ($\epsilon << 1$), one can ignore the second term in the right-hand side,
and the contribution to ZEDOS from the nested part of the Fermi surface
$\nu_{nest}(0)$ is written as
\begin{eqnarray}
\nu_{nest}(0) \propto \frac{1}{\sqrt{1+\frac{2\Lambda}{\pi^2}|\Delta_0|^2\bigl(\frac{|\Phi(\mathbf{k}_F)|}{v_F\sin\phi_0}\bigr)^2}}\nonumber\\ 
 \sim \frac{\pi v_F}{\sqrt{2\Lambda}|\Delta_0\Phi(\mathbf{k}_F)|}|\sin\phi_0|, \label{cusporigin}
\end{eqnarray}
for small $\phi_0$. Hence, FODOS has cusp-like minima at $\phi_0=0$,
i.e., when a magnetic field is applied perpendicular to the nested
parts.

\begin{figure}
\includegraphics[width=0.5\textwidth,scale=0.1]{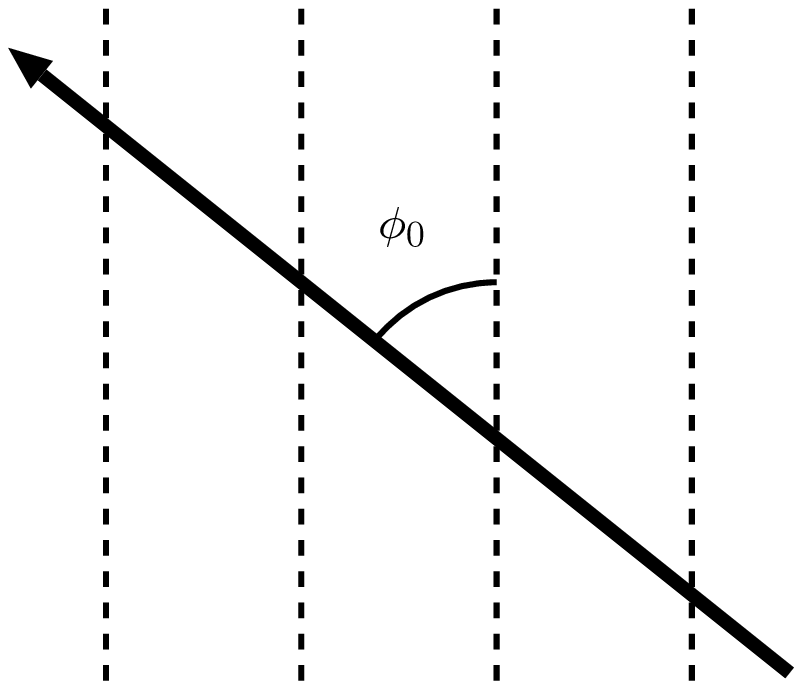}
\includegraphics[width=0.5\textwidth]{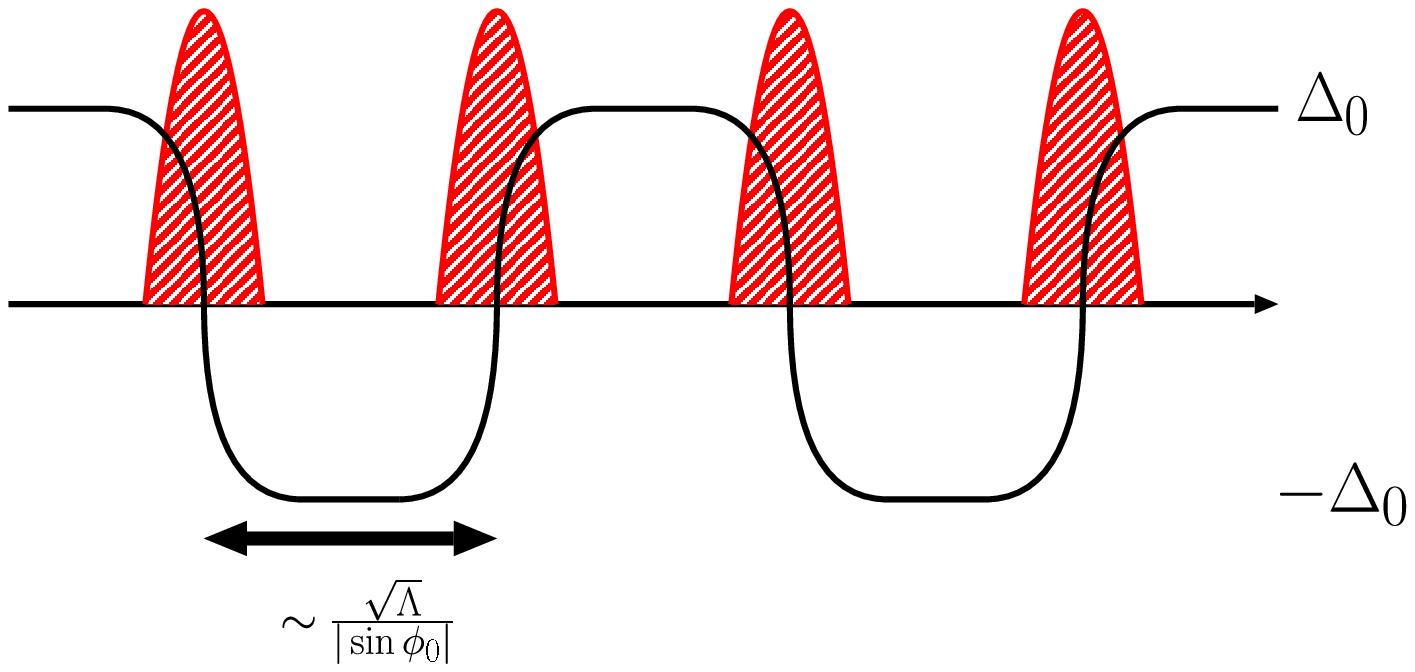}
\caption{\label{corestate} (upper)A schematic picture of a trajectory running
 at an angle $\phi_0$ with magnetic field. Vortex cores are shown with
 dashed lines. (lower)Andreev bound states formed on the trajectory.}
\end{figure}

The $|\sin\phi_0|$ dependence in eq. (\ref{cusporigin}) allows one a
simple physical interpretation. Suppose a quasiclassical
trajectory\cite{serene,kopnin,schopohl2} of nested Fermi surface, which is running in the
direction at an angle $\phi_0$ with magnetic field, and passing near
vortex cores (fig. \ref{corestate}). On this trajectory, phase of order parameter changes
drastically near vortex cores, leading to formation of Andreev bound
states. Since the vortex cores are separated by the distance
$\sim\frac{\sqrt{\Lambda}}{|\sin\phi_0|}$ on the trajectory, the
density of Andreev bound states is proportional to
$\frac{|\sin\phi_0|}{\sqrt{\Lambda}}$, hence make a cusp-like
singularity in FODOS. This brief discussion reveals that the cusp-like
singularity is caused by the core states, which are not properly taken into
account in the Doppler-shift method.

Here, we discuss the basic properties of YNi$_2$B$_2$C.
YNi$_2$B$_2$C has three bands (17th,18th,19th) crossing Fermi
energy\cite{singh, lee, dugdale}. The 17th band, which mainly consists
of Ni(4d) orbits, forms an open Fermi surface in the c-axis direction,
hence has a quasi-two-dimensional nature\cite{singh, lee, dugdale}. It
is reported that the corresponding Fermi surface of LuNi$_2$B$_2$C has a
nested part, with nesting vectors $\mathbf{q}\sim (0.56\times2\pi)\mathbf{e}_a,
(0.56\times2\pi)\mathbf{e}_b$\cite{dugdale}. Kohn anomaly in phonon
spectrum\cite{dervenagas,kawano,stassis} provides another evidence of
the nesting. On the other hand, the 18th and the 19th bands form closed
Fermi surface in the c-axis direction, hence have three-dimensional
tendency. As to the gap structure, existence of nodes have been
confirmed by specific heat\cite{nohara} and ultra-sonic
attenuation\cite{watanabe2} etc. 

In order to discuss the features of YNi$_2$B$_2$C, we study the
 following two-band model. Our model consists of ($\alpha$) a partly nested
quasi-two-dimensional Fermi surface as shown in fig. \ref{necylinder},
and ($\beta$) a three-dimensional spherical Fermi surface. We suppose the
normal-state density of states of these two Fermi surfaces are the same.
We set $\zeta=v/u=0.137$ for the first Fermi surface. Then, about 4\% of
the entire Fermi surfaces are nested, in agreement with
ref. \onlinecite{dugdale}, where the proportion of the nested part is reported
to be $4.4\pm 0.5$\%. As to the gap structure, we assume
isotropic gap for the quasi-two-dimensional Fermi surface, and 
the following two cases for the spherical Fermi surface, with different nodal position:
(i)$\Phi(\mathbf{k}_F)=\sqrt{\frac{15}{4}}\sin^2\theta\sin 2\phi$ and
(ii) $\Phi(\mathbf{k}_F)=\sqrt{\frac{15}{4}}\sin^2\theta\cos 2\phi$.
As to the interaction potential $V^{\alpha\beta}$ in
eq. (\ref{mpeschgapeq}), we assume
 $V^{\alpha\beta}/V^{\alpha\alpha}=0.5$ and
 $V^{\beta\beta}/V^{\alpha\alpha}=3.0$. Here, we assume
spherical Fermi surface has the larger superconducting instability, in
order to make superconducting coherence lengths of this system isotropic\cite{takagi}.

\begin{figure}
\includegraphics[width=0.5\textwidth]{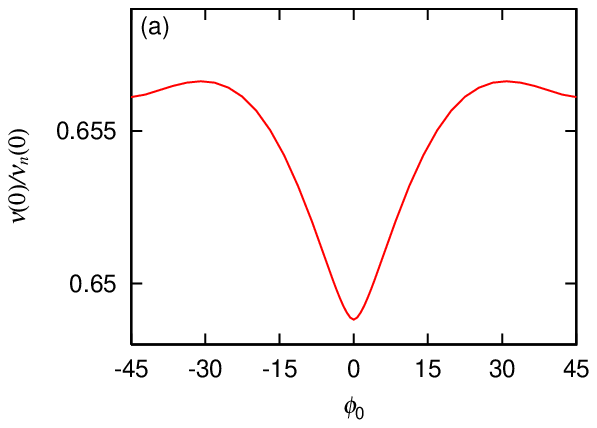}
\includegraphics[width=0.5\textwidth]{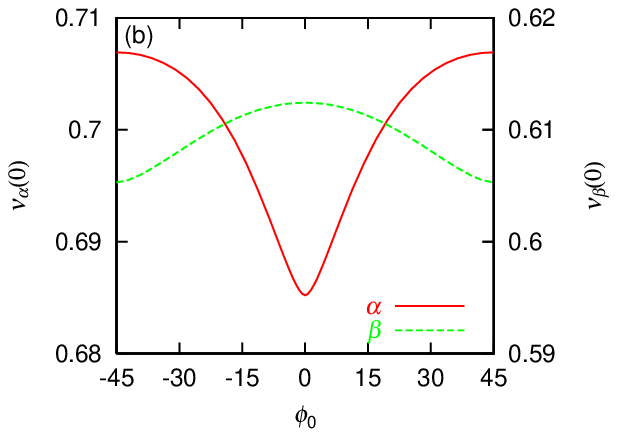}
\caption{\label{bolofodos1}(a)FODOS of the two-band model. Gap function
 (i) is adopted for the nodal structure of the spherical Fermi
 surface. (b)Contribution to FODOS from each Fermi surface.}
\end{figure}
\begin{figure}
\includegraphics[width=0.5\textwidth]{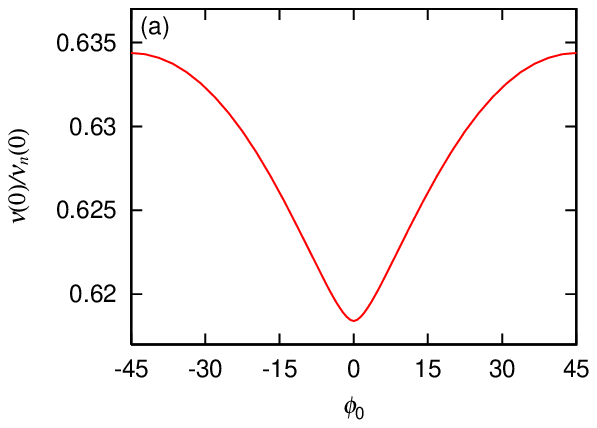}
\includegraphics[width=0.5\textwidth]{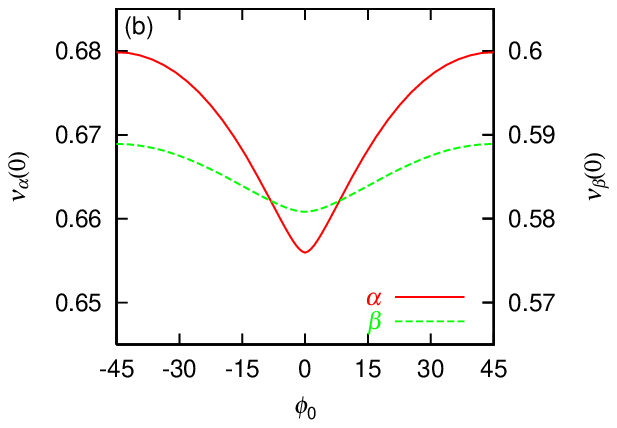}
\caption{\label{bolofodos2}(a)FODOS of the two-band model. Gap function
 (ii) is adopted for the nodal structure of the spherical Fermi
 surface. (b)Contribution to FODOS from each Fermi surface.}
\end{figure}
\begin{figure}
\includegraphics[width=0.5\textwidth]{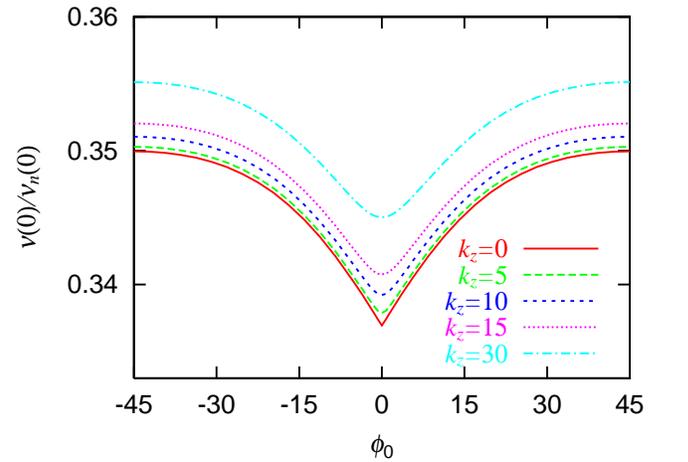}
\caption{\label{bolofodos_div}Contribution to FODOS from several $k_z$
 of the quasi-two-dimensional Fermi surface. This figure should be
 compared with fig. \ref{fermi_div}(b)}
\end{figure}

In fig. \ref{bolofodos1}(a) and fig. \ref{bolofodos2}(a), we give FODOS
of this model. Calculation is done at
$H=0.1H_{c2}(\phi_0=0^{\circ})$. One can find a clear cusp-like minima in both
cases. Furthermore, it is worth noting that FODOS has minima for $H\parallel a$($\phi_0=0$),
irrespective of the nodal position. In fig. \ref{bolofodos1}(b) and
fig. \ref{bolofodos2}(b), we show contribution to FODOS from each Fermi
surface. One can find that the FODOS arising from the
quasi-two-dimensional Fermi surface is larger and dominate the
contribution from the spherical Fermi surface. It is surprising that
only tiny anisotropy of Fermi surface destroys the Doppler-shift
predominant region below $H=0.1H_{c2}(\phi_0=0^{\circ})$. It makes a contrast with the single
band case in section \ref{inplane} where as much as 40\% of Fermi
surface has to be nested to suppress the Doppler-shift predominant
region to $H\sim 0.05H_{c2}$. 

One can understand the origin of the enhancement of the contribution from the
quasi-two-dimensional Fermi surface, on the basis of the scenario in
section \ref{isotropic}. In section \ref{isotropic}, we noted that largest contribution to FODOS
comes from the part of Fermi surface where effective Fermi velocity
$\mathbf{v}_{F\perp}$ is nearly parallel to the field-rotational plane.
In the case of cylindrical Fermi surface in section \ref{isotropic},
such parts are limited to $k_z\sim 0$ due to
anisotropic coherence lengths $\delta\sim\epsilon$. However, in the
two-band model in this section, the superconducting coherence lengths
are isotropic ($\delta\sim 1$), hence $\mathbf{v}_{F\perp}$ is almost parallel to the
field-rotational plane all over the quasi-two-dimensional Fermi surface. Therefore,
contribution to FODOS come from all $k_z$ and becomes much larger than the
contribution from the nodal structure on the spherical Fermi surface. We
plot contribution to FODOS from several $k_z$ in
fig. \ref{bolofodos_div}. In this figure, one can find that all $k_z$
contribute to FODOS equally.

The cusp-like singularity shown here is basically owing to the
co-existence of two-dimensional and three-dimensional Fermi surfaces and
to the nesting nature of two-dimensional Fermi surface. This example
shows that many cares are needed for an analysis of experimental data
in the multi-band system. Then, anisotropy of quasi-two-dimensional
Fermi surface is enhanced and invalidate the simple argument
based on the Doppler-shift methods.

\section{Conclusions}
\label{conclusion}
We study the influence of superconducting gap and Fermi surface
structures on the field-orientational dependence of ZEDOS. 
First, we studied the crossover behavior of ZEDOS, by taking account of
the four-fold oscillation of $\Delta_0$, and found that there exists a crossover
magnetic field $H^*$ so that $\nu_{anti}(0) > \nu_{node}(0)$ for $H <
H^*$ , while $\nu_{node}(0) > \nu_{anti}(0)$ for $H > H^*$, consistent
with our previous analyses\cite{udagawa}.

Next, we have investigated the roles of gap structure and nature of
Fermi surface. The crossover behavior and FODOS at the Doppler-shift
predominant region have been discussed. We found that there are no
significant differences in neither the value of $H^*$ nor the shape of
FODOS between the spherical Fermi surface and the
cylindrical Fermi surface, in spite of the difference in the
dimensionality of Fermi surface. We also found that when quadratic point
nodes exist, $H^*$ is lowered than the line node cases and FODOS has
broader minima. These results can be coherently understood by
recognizing that contribution to FODOS mostly comes from the part on the
Fermi surface where effective Fermi velocity is nearly parallel to the
field-rotational plane.

Next, we examine the effects of in-plane anisotropy of Fermi surface. We
showed that the Doppler-shift predominant region may be suppressed to
lower fields by the anisotropy of Fermi surface. This tendency becomes
especially strong in a multi-band superconductor which has nearly isotropic
coherence lengths. Therefore, some cares are necessary for analyzing the
field-rotational experiments for YNi$_2$B$_2$C and CeCoIn$_5$. On the
other hand, the field-rotational experiment and its interpretation based
on the Doppler-shift criterion are powerful tool for such material as
Sr$_2$RuO$_4$, Na$_{0.35}$CoO$_2$1.3H$_2$O\cite{takada} and organic superconductors, which are well described by a
single Fermi surface or have plural Fermi surfaces
with the same dimensionality.

Finally, we propose a novel mechanism of the cusp-like minima found in
YNi$_2$B$_2$C, in terms of Andreev bound states. The cusp-like minima 
are attributed to nesting in quasi-two-dimensional Fermi surface and the
isotropy of superconducting coherence lengths.

We are grateful to Y. Matsuda, T. Sakakibara, Y. Kato, K. Izawa, H. Kusunose and
T. Watanabe for fruitful discussions. We would also like to thank
K. Kamata for providing her master thesis on the field-rotational
experiments of YNi$_2$B$_2$C.

\end{document}